\documentclass[twocolumn,showpacs,prl,aps]{revtex4-1}
\usepackage{bm,color,amsmath,bbold,bbm,amssymb,mathrsfs,latexsym,graphicx,psfrag}

\newcommand{\bs}[1]{\boldsymbol{#1}}
\newcommand{\be}{\begin{equation}}
\newcommand{\ee}{\end{equation}}
\newcommand{\bea}{\begin{eqnarray}}
\newcommand{\eea}{\end{eqnarray}}

\renewcommand{\phi}{\varphi}
\renewcommand{\epsilon}{\varepsilon}
\renewcommand{\vec}[1]{{\bf #1}}
\newcommand{\pdagger}{\phantom{\dagger}}

\def\nn{\nonumber\\}

\usepackage{soul}
\sethlcolor{green}
\setstcolor{red}

\begin{document}

\title{Anisotropic chiral $d+id$ superconductivity in $\text{Na}_x\text{CoO}_2\cdot y \text{H}_2\text{O}$}

\author{Maximilian L. Kiesel${}^1$} 
\author{Christian Platt${}^1$} 
\author{Werner Hanke${}^1$} 
\author{Ronny Thomale${}^{2}$}

\affiliation{${}^1$Institute for Theoretical
  Physics, University of W\"urzburg, Am Hubland, D
  97074 W\"urzburg} 
\affiliation{${}^2$ Institut de th\'eorie des ph\'enom\`enes physiques, \'Ecole Polytechnique F\'ed\'erale de Lausanne (EPFL), CH-1015 Lausanne}
\date{\today}

\begin{abstract}
Since its discovery, the superconducting phase in water-intercalated sodium cobaltates $\text{Na}_x\text{CoO}_2~\cdot~y\text{H}_2\text{O}$ ($x\sim 0.3$, $y\sim 1.3$) has posed fundamental challenges in terms of experimental investigation and theoretical understanding. By a combined dynamical mean-field and renormalization group approach, we find an anisotropic chiral $d+id$ wave state as a consequence of multi-orbital effects, Fermi surface topology, and magnetic fluctuations. It naturally explains the singlet property and close-to-nodal gap features of the superconducting phase as indicated by experiments.    
\end{abstract}


\maketitle

{\it Introduction.} Initiated by the discovery of the cuprates, the search for new materials exhibiting unconventional superconductivity has become one of the major branches in condensed matter physics~\cite{sigrist-91rmp239}. A particularly exciting idea  is the concept of chiral superconductivity, where the Cooper pair condensate breaks parity and time reversal symmetry and gives rise to interesting edge mode phenomena of the bulk-gapped superconductor. As one way to accomplish such a scenario, the lattice can act as a custodial symmetry to ensure the exact, or, in the presence of disorder and nematic fluctuations, approximate degeneracy of different superconducting instabilities. In such a case, the degeneracy is linked to higher dimensional irreducible representations of the lattice symmetry group, and a chiral superposition of superconducting states can be energetically favorable below T$_c$. Unfortunately, for the square lattice and its $C_{4v}$ group, there is no such representation for singlet Cooper pairs, which in the majority of materials is found to be the generic sector for superconductivity. This, however, changes for hexagonal systems, where the $E_2$ representation of the $C_{6v}$ lattice symmetry group implies the degeneracy of the $d_{x^2-y^2}$ and $d_{xy}$ wave state at the instability level~\cite{kiesel-12prb020507}, which can yield a chiral $d+id$ singlet superconductor.

In many respects, the water-intercalated sodium cobaltates $\text{Na}_x\text{CoO}_2\cdot y \text{H}_2\text{O}$, with a superconducting dome for $x\sim 0.3$, $y\sim 1.3$ at T$_c=4.5$K~\cite{takada-03n53}, have been interpreted as the natural generalization of a square lattice of copper oxide in the high-T$_c$ cuprates to a triangular lattice of cobalt oxide: The electronic structure can be assumed effectively two-dimensional due to the intercalation, and superconductivity emerges as a function of sufficient Na doping in proximity to magnetic phases. The experimental evidence, however, remained inconclusive for a significant amount of time, which did not allow to draw substantiated conclusions on the nature of the order parameter~\cite{mazin-05np91}. For example, previous ambiguous indications from Knight-shift measurements for polycrystalline samples have only later been clarified by single crystal measurements~\cite{zheng-06prb180503}, which showed the singlet property of the superconducting phase. Similarly, the nodal character of the order parameter has remained a contentious issue. Early $\mu$SR~\cite{kanigel-04prl257007} as well as magnetic penetration depth measurements~\cite{fujimoto-04prl047004}, have shown evidence against a homogeneous gap and have been interpreted in favor of line nodes. By contrast, the latest specific heat studies~\cite{oeschler-08prb054528} advocate a two gap scenario for the cobaltates with one comparably small and another slightly larger gap to fit the data. This reminds us of similar discussions for the iron pnictides, where it is likewise complicated to distinguish a nodal from a strongly anisotropic gap~\cite{thomale-11prl187003}.

In this Letter, we develop a microscopic theory for the nature of superconductivity in the sodium cobaltates which is consistent with the experimental findings. Inspired by the resemblance to the cuprates, the earliest theoretical proposals employed a phenomenological RVB theory for the cobaltates~\cite{baskaran03prl097003} supplemented by slave boson mean-field calculations~\cite{wang-04prb092504}. A major challenge from the beginning has been the choice of an adequate low energy kinetic theory for the problem: While ARPES measurements only observe one Fermi pocket centered around $\Gamma$ in the hexagonal Brillouin zone~\cite{hasan-04prl246402,yang-04prl246403,yang-05prl146401}, band structure calculations indicate the presence of additional $e'_g$ pockets~\cite{singh00prb13397}. 
The absence of the $e'_g$ pockets in the experimental cobaltate scenario has been assigned to surface effects~\cite{pillay-08prl246808}, disorder~\cite{singh-06prl016404}, and electronic correlations~\cite{zhou-05prl206401,marianetti-07prl246404,wang-08prl066403}. It suggests that whatever the microscopic theory for superconductivity in the cobaltates may be, it should involve a low-energy kinetic theory which explains the experimental evidence from a single-pocket scenario. Such an effective model has been developed by a combined dynamical mean-field and cluster approximation approach by Bourgeois et al.~\cite{bourgeois-09prl066402,bourgeois-07prb174518}, which is the starting point of our investigations. Employing multi-orbital functional renormalization group (fRG)~\cite{wang-09prl047005,thomale-09prb180505,thomale-11prl187003,platt-11prb235121,phtinprep} to obtain an effective interaction profile for this model, we find a rich phase diagram for the sodium cobaltates with an anisotropic $d+id$-phase in the relevant doping regime. The strong anisotropy of the superconducting gap can explain the experimental evidence; it follows from the interplay of multi-orbital hybridization, Fermi surface topology, and frustrated magnetic fluctuations in the sodium cobaltates.

{\it Cobaltate effective kinetic model.}  
  \begin{figure}[t]
    \begin{minipage}[l]{0.99\linewidth}
      \includegraphics[width=\linewidth]{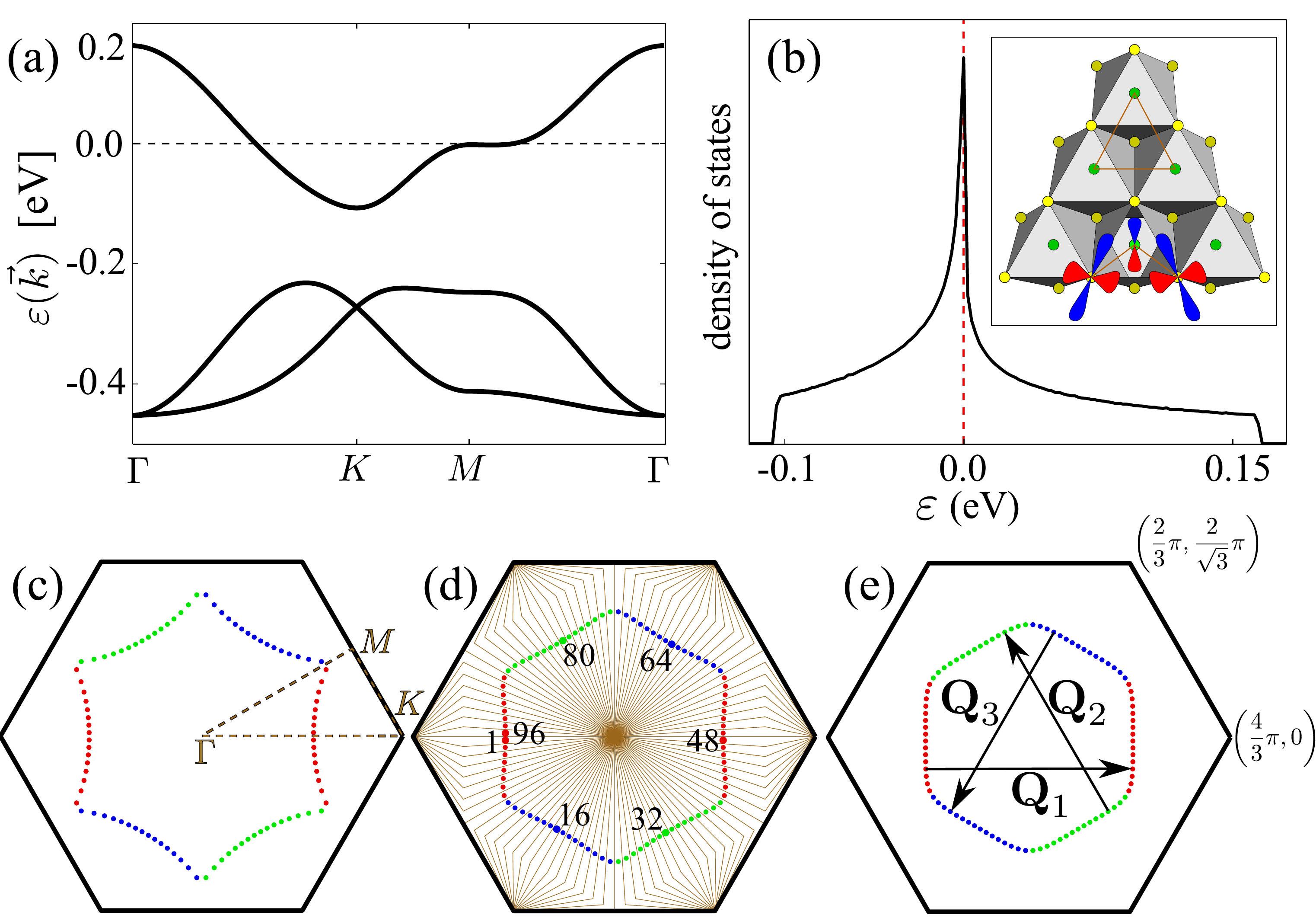}
    \end{minipage}
    \caption{(Color online). (a) Effective band structure resulting from~\eqref{eq:hamiltonian} with $t=0.1$eV, $t'=-0.02$eV and $D=0.105$eV. (b)~The van Hove singularity is visible in the density of states (inset: crystal structure of the $\text{CoO}_2$ layers). The Fermi surface is shown at $x=0.1$ (c), $x=0.2$ (d), and $x=0.3$ (e). The Fermi surface colors indicate the dominant orbital weights. (d) sketches the division of the Brillouin zone into 96 patches used in fRG, (e) depicts the nesting vectors $\vec{Q}_N$.} 
    \label{fig:band_structure}
  \end{figure}
Following the work by Bourgeois et al.~\cite{bourgeois-09prl066402,bourgeois-07prb174518}, all $\text{Co}_{3d}$ and $\text{O}_{2p}$ orbitals are taken into account in a finite cluster calculation which is then mapped to a three-orbital model and fitted against X-ray absorption and ARPES data~\cite{bourgeois-07prb174518}. The resulting effective model is obtained from dynamical mean-field theory calculations, which adequately take into account the self-energy effects at a single-particle level. It exhibits strongly hybridized orbitals formed by an effective $t_{2g}$ manifold  $(\tilde{d}_{xy},\tilde{d}_{yz},\tilde{d}_{zx})$ per site on the triangular Co super lattice, a finding which is also consistent with ARPES polarization measurements~\cite{qian-06prl186405}. The Hamiltonian reads
\begin{widetext}
\begin{eqnarray}
    H_{\text{eff}} &=& \sum \limits_{\langle i,j\rangle,\alpha\beta,\sigma}  \left(\left(t + t'\delta_{\alpha\beta} + D\delta_{ij} \right) \hat{c}_{i\alpha\sigma}^{\dagger} \hat{c}_{j\beta\sigma}^{\pdagger} + \text{h.c.}\right) 
    + \mu \sum \limits_{i,\alpha,\sigma} \hat{n}_{i\alpha\sigma} + U_1 \sum \limits_{i,\alpha} \hat{n}_{i\alpha\uparrow} \hat{n}_{i\alpha\downarrow} \nn 
  &&  + \frac{1}{2} \sum \limits_{i,\alpha\neq\beta} \left( U_2 \sum \limits_{\sigma,\nu} \hat{n}_{i\alpha\sigma} \hat{n}_{i\beta\sigma\nu} \right. 
    + \left. J_{\text{H}} \sum \limits_{\sigma,\nu} \hat{c}_{i\alpha\sigma}^{\dagger} \hat{c}_{i\beta\nu}^{\dagger} \hat{c}_{i\alpha\nu}^{\pdagger} \hat{c}_{i\beta\sigma}^{\pdagger} + J_{\text{P}} \hat{c}_{i\alpha\uparrow}^{\dagger} \hat{c}_{i\alpha\downarrow}^{\dagger} \hat{c}_{i\beta\uparrow}^{\pdagger} \hat{c}_{i\beta\downarrow}^{\pdagger} \right) , \label{eq:hamiltonian}
\end{eqnarray}
\end{widetext}
 where $ \hat{n}_{i\alpha\sigma}= \hat{c}_{i\alpha\sigma}^{\dagger} \hat{c}_{i\alpha\sigma}^{\pdagger}$, and $\hat{c}_{i\alpha\sigma}^{\dagger}$ denotes the electron creation operator of spin $\sigma={\uparrow,\downarrow}$ in orbital $\alpha=1,2,3$ at site $i$. $t$ represents the hopping mediated by $\text{O}_{2p_\pi}$, $t'$ corresponds to a direct Co-Co-hopping, $D$ is the crystal-field splitting, and $\mu$ the chemical potential. We set $t = 0.1$eV, $t' = -0.02$eV, and $D=0.105$eV~\cite{bourgeois-09prl066402,bourgeois-07prb174518}. The bandwidth of the effective model is $\sim0.6$eV. This is a factor $3$ smaller than LDA calculations predict ($1.6$eV~\cite{singh00prb13397} or $2.0$eV~\cite{Lee-04prb045104}).
This effective three-band model resulting from~\eqref{eq:hamiltonian} yields one band intersecting the Fermi level (Fig.\ref{fig:band_structure}a). A van Hove singularity occurs at a doping level of $x\approx0.09$ (Fig.\ref{fig:band_structure}b). The Fermi surface contains one hole pocket around $\Gamma$, i.e. the center of the Brillouin zone (Fig.\ref{fig:band_structure}c). All three hybridized orbitals contribute to the Fermi surface and each has two antipodal dominant regions, indicated by red/green/blue dots in Fig.\ref{fig:band_structure}c-e, corresponding to the doping $x~=~\{0.1,0.2,0.3\}$. At $x\approx0.28$, the nesting of the Fermi surface is optimal, with three nesting wave-vectors
$\vec{Q}_N \approx \pi \{ \left( \sqrt{2},0 \right), ( -\frac{1}{\sqrt{2}},\sqrt{\frac{3}{2}} ) , ( -\frac{1}{\sqrt{2}},-\sqrt{\frac{3}{2}}) \}$ (Fig.~\ref{fig:band_structure}e).
The interaction part of~\eqref{eq:hamiltonian} features intra-orbital Coulomb interaction $U_1$, inter-orbital Coulomb interaction $U_2$, Hund's rule coupling $J_{\text{H}}$, and pair hopping $J_{\text{P}}$. 
As the bandwidth in this effective model is smaller than in bare LDA calculations, the interactions strengths from bare ab-initio calculations likewise have to be regularized. We set $U_1=0.37$eV, $U_2=0.25$eV, and $J_{\text{H}}=J_{\text{P}}=0.07$eV.

We now proceed by investigating the Fermi surface instabilities of~\eqref{eq:hamiltonian} through fRG, where we use the effective multi-orbital band structure as the initial starting point~\cite{wang-09prl047005,thomale-09prb180505,thomale-11prl187003,platt-11prb235121,phtinprep}. Through renormalization, we obtain an effective low-energy theory of the scattering vertex which exhibits superconductivity. The pairing 2-particle vertex $V^{\text{SC}}(\bs{k},\bs{q})$ is then decomposed into eigenmode contributions which correspond to the different superconducting form factors $V^{\text{SC}}(\bs{k},\bs{q})=\sum_i c_i^{\text{SC}} f^i(\bs{k}) f^i(\bs{p})$. $c_i^{\text{SC}}$ signals the strength of the instability and hence allows to identify the superconducting phase adopted by the system. We employ multi-orbital temperature-flow fRG~\cite{platt-11prb235121} to take into account the interplay between ferromagnetic and antiferromagnetic fluctuations as well as the multi-orbital character of the sodium cobaltates. Note, that we must avoid the double counting of self-energy effects that have already been included to obtain this band structure. As such, we intentionally do not take into account self-energy effects at the single-particle level which emerge during the RG flow. In total, our procedure is still an approximation, in the sense that we first renormalize the single-particle level via a DMFT approach and then separately investigate the renormalization of the scattering vertex through fRG which gives rise to Fermi surface instabilities. This, however, is justified because the scattering vertex evolution under RG is only significant in the immediate vicinity of the Fermi level.

\begin{figure}[t]
    \begin{minipage}[l]{0.99\linewidth}
      \includegraphics[width=\linewidth]{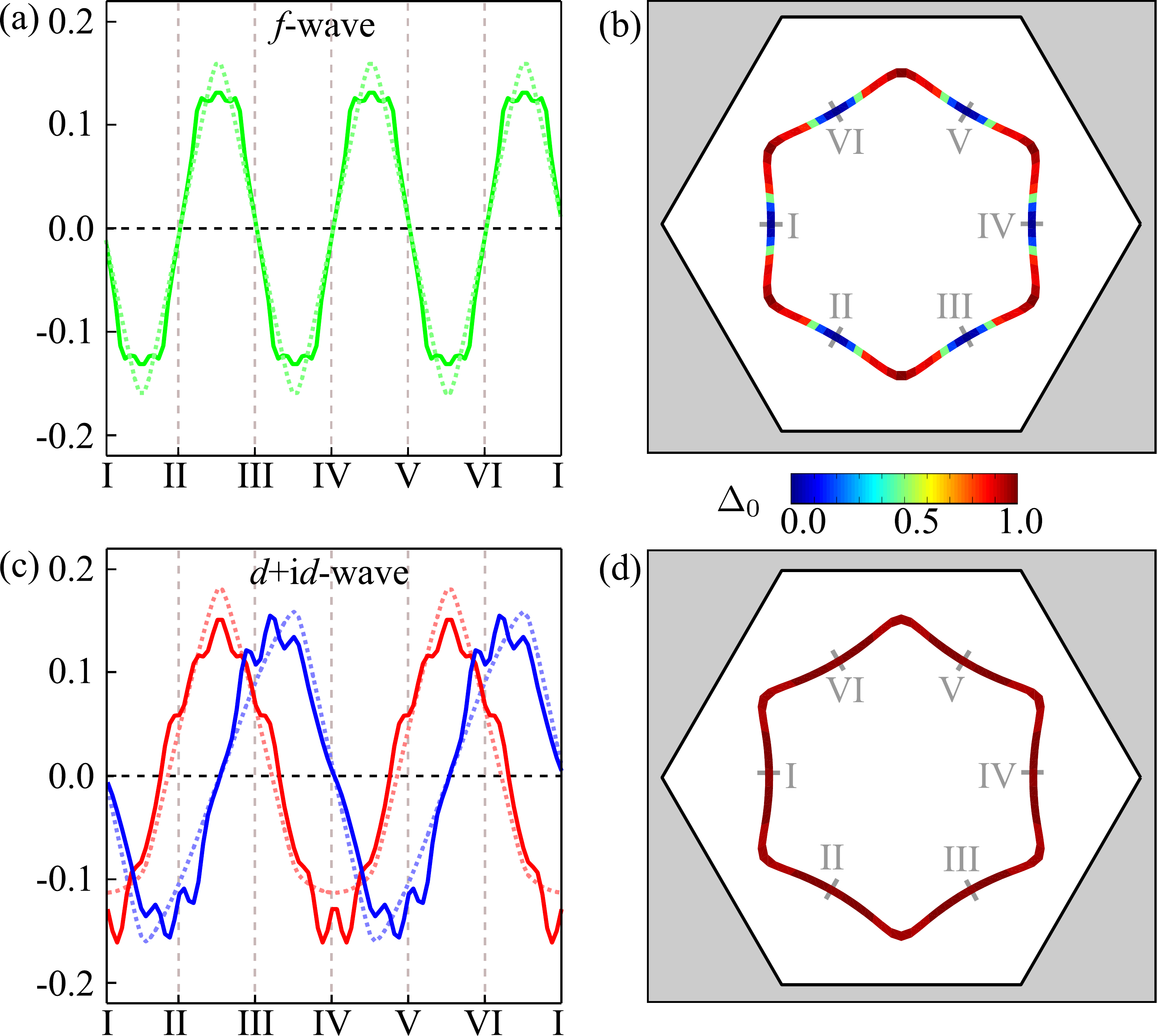}
    \end{minipage}
    \caption{(Color online). Superconducting form factors from fRG mean-field decoupling of pairing channels (solid lines) compared to analytic leading harmonic solutions (dashed lines): (a,b) $f$-wave, (c,d) $d+id$-wave. The representative parameters are (a,b) $U_1/U_2=1.4$, $x=0.18$ and (c,d) $U_1/U_2=1.0$, $x=0.14$. (b) and (d) show a color plot of the gap size $\Delta_0$ along the Fermi surface (Eq.~\ref{eq:gap}).}
    \label{fig:form_factors}
  \end{figure}

\begin{figure}[t]
    \begin{minipage}[l]{0.99\linewidth}
      \includegraphics[width=\linewidth]{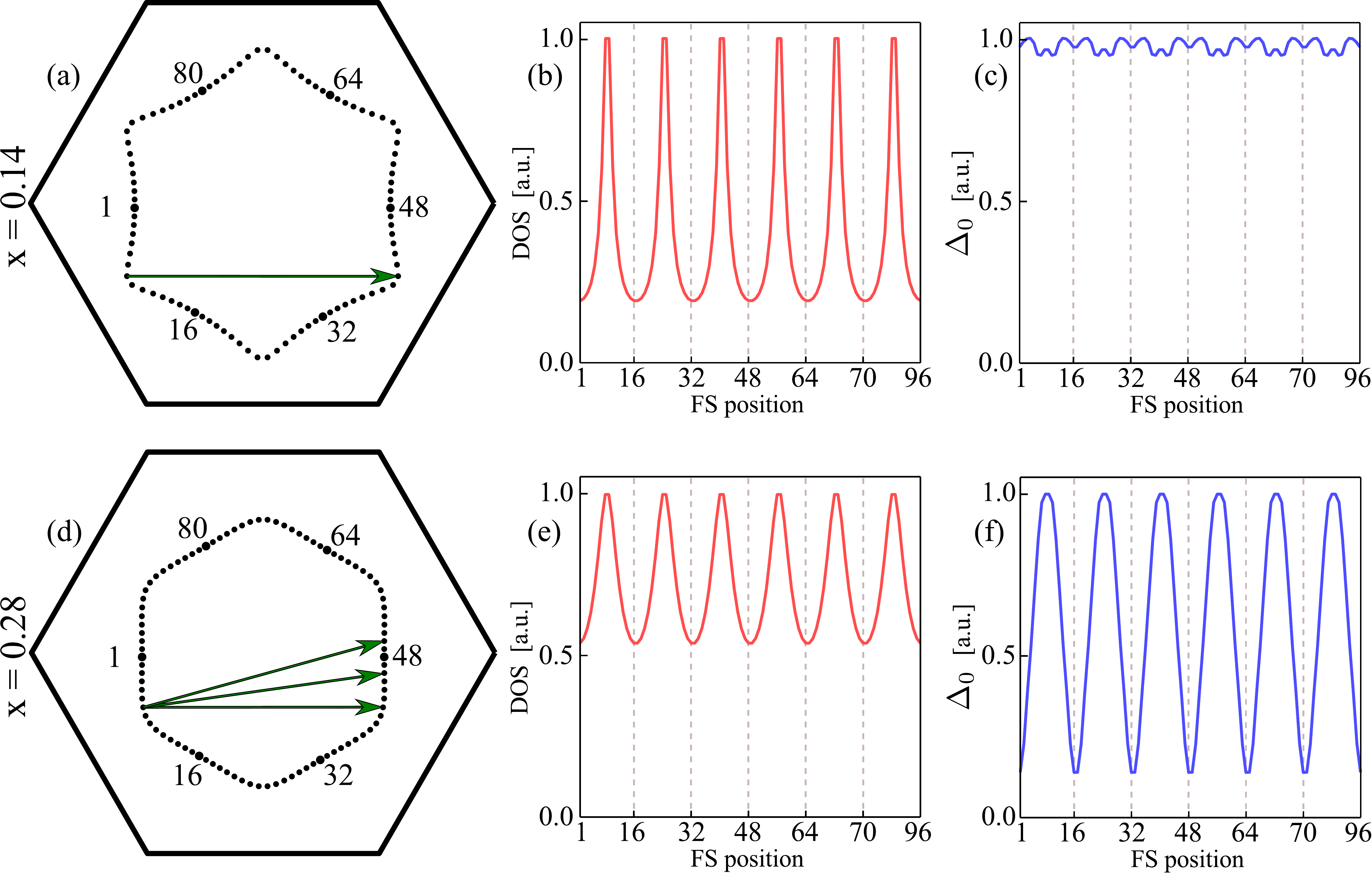}
    \end{minipage}
    \caption{(Color online) Change of gap anisotropy in the $d+id$ superconducting phase as a function of doping. (a-c) $x=0.14$. The dominant density of states  is strongly peaked at the warped edges of the Fermi surface and singles out specific particle hole scattering channels which result in a homogeneous gap $\Delta_0$. (d-f) $x=0.28$. Various scattering channels are enhanced due to reduced Fermi surface warping and more homogenous distribution of density of states, yielding a strongly anisotropic gap.} 
    \label{fig:anisotropy}
  \end{figure}

{\it Phase diagram.}
\begin{figure}[t]
\begin{minipage}[l]{0.99\linewidth}      \includegraphics[width=\linewidth]{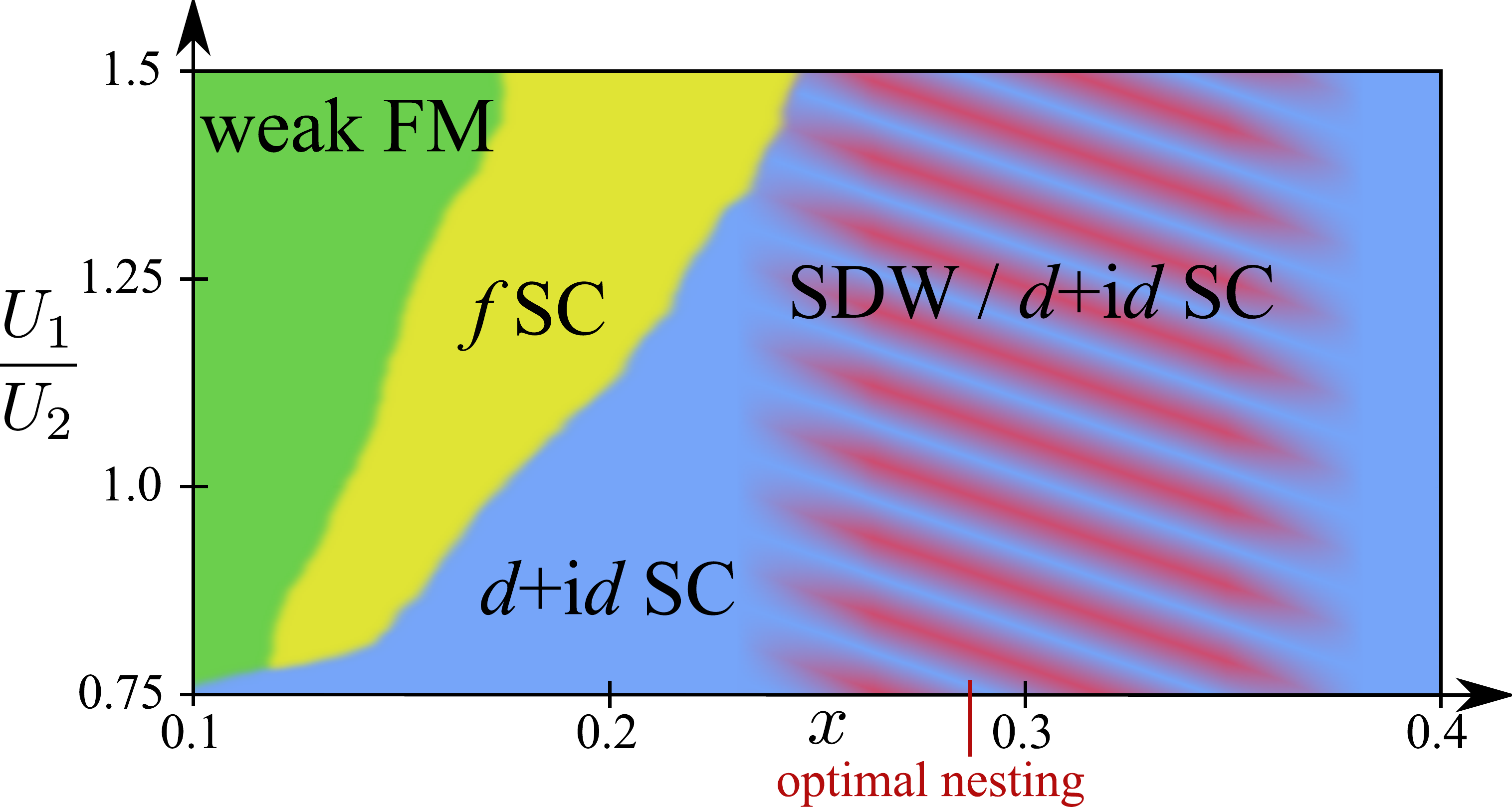}
\end{minipage}
    \caption{(Color online). Phase diagram of the model~\eqref{eq:hamiltonian} as function of doping $x$ and $U_1/U_2$. There are four phases: $d+id$-wave superconductivity ($d+id$ SC, blue), weak ferromagnetism (weak FM, green), $f$-wave superconductivity ($f$ SC, yellow), and a phase with competitive spin-density wave and $d+id$-wave superconductivity (SDW / $d+id$ SC, purple and blue shaded).}
    \label{fig:phase_diagram}
  \end{figure}
From an itinerant viewpoint of Fermi surface instabilities, an important feature of~\eqref{eq:hamiltonian} is the different doping location of the van Hove singularity ($x=0.09$) and the optimally nested Fermi surface ($x=0.28$). Note that both locations are coincident and much less revealing for a triangular lattice tight-binding model with only nearest neighbor hopping~\cite{honerkamp03prb104510}. Accordingly, for small doping, the phases are determined by the large density of states at the Fermi level combined with rather weak nesting corresponding to finite momentum transfer, which in total promotes dominant zero momentum particle hole scattering (labeled weak ferromagnetism (FM) in Fig.~\ref{fig:phase_diagram}). With increased doping, it depends on the ratio of $U_1/U_2$ whether the system favors triplet $f$-wave or singlet $d+id$-wave superconductivity. For the former, the system exhibits nodes along the Fermi surface and follows the gap function  $f \left( \bs{k} \right) = \sin(k_y) - 2 \cos(\frac{\sqrt3 k_x}{2}) \sin(\frac{k_y}{2}) $ (Fig.~\ref{fig:form_factors}a,b). $f$-wave is preferred for enhanced $U_1/U_2$, as it is seeded by spin alignment stemming from ferromagnetic fluctuations, which are reduced by $U_2$.
In the case of preferred $d$-wave superconductivity, we find two degenerate instabilities associated with the form factors depicted in Fig.~\ref{fig:form_factors}c, which relate to the leading harmonics $d_{x^2-y^2} \left( \bs{k} \right) = 2 \cos(k_x) - \cos(\frac{k_x - \sqrt3 k_y}{2}) - \cos(\frac{k_x + \sqrt3 k_y}{2}) $ and
$ d_{xy} \left( \bs{k} \right) = \cos(\frac{k_x + \sqrt3 k_y}{2}) - \cos(\frac{k_x - \sqrt3 k_y}{2}) $. (Note that throughout parameter space, we always find higher harmonic contributions in the $d$-wave sector~\cite{zhou-08prl217002} to be irrelevant.) The system could generically form any linear combination $ d_1+e^{i\theta}d_2$ of both $d$-wave solutions which must be degenerate at the instability level as protected by lattice symmetry. A mean-field decoupling in the SC pairing channel and minimization of the free energy as a function of the superposition parameter, can be rephrased by satisfying the self-consistent gap equation~\cite{kiesel-12prb020507}
  \begin{equation}
    \Delta_{\bs{q}}=-1/N\sum_{\bs{k}}V^{\text{SC}}(\bs{k},\bs{q})\frac{\Delta_{\bs{k}}}{2 E(\bs{k})} \text{tanh} \left( \frac{E(\bs{k})}{2T}\right). \label{eq:gap}
  \end{equation}
  We always find $d+id$ to be the energetically preferred combination. This is rather generic in a situation of degenerate nodal SC order parameters, since such a combination allows the system to avoid nodes in the gap function (Fig.~\ref{fig:form_factors}d) and maximizes condensation energy. Note, however, that the relative energy gain between a $d+id$ state and a different possible solution such as nodal single $d_{x^2-y^2}$ varies significantly depending on the microscopic setup. For example, the condensation energy gain from $d+id$ for lower doping in Fig.~\ref{fig:anisotropy}c as compared to a single $d$ wave solution will be higher than for larger doping in Fig.~\ref{fig:anisotropy}f.
 
{\it Anisotropic regime at $x \sim  0.3$.} The effective model in~\eqref{eq:hamiltonian} is quantitatively most accurate in the doping regime of the superconducting dome of the cobaltates. There, we find $d+id$ superconductivity with strong SDW fluctuation background (Fig.~\ref{fig:phase_diagram}). The enhancement of magnetic fluctuations is also observed in experiment, anticipating the metal insulator transition regime at $x\sim 0.5$. From the viewpoint of Fermi surface topology, this is due to improved nesting conditions for a larger part of Fermi level density of states (Fig.~\ref{fig:band_structure}e). As the density is also more homogeneously distributed along the parallel sides of the hexagonal Fermi surface, however, this leads to an increased bandwidth of enhanced particle-hole channels in the RG flow, which eventually yields an enhanced anisotropy in the seeded pairing channel which gives rise to $d+id$ superconductivity.
This trend for increasing doping $x$ is illustrated in Fig.~\ref{fig:anisotropy}. At lower doping, the more warped Fermi surface clearly singles out the particle hole channel according to scattering between the Fermi surface edges, which also possess the dominant fraction of density of states at the Fermi level (Fig.~\ref{fig:anisotropy}a-c). As doping is increased, the reduced warping allows the effectively one-dimensional parallel sections of the Fermi surface to drive more particle-hole channels. The result is an evolution of the seeded superconducting phase from a homogeneous to an extremely anisotropic $d+id$ gap (Fig.~\ref{fig:anisotropy}d-f).

\begin{figure}[t]
    \begin{minipage}[l]{0.99\linewidth}
      \includegraphics[width=\linewidth]{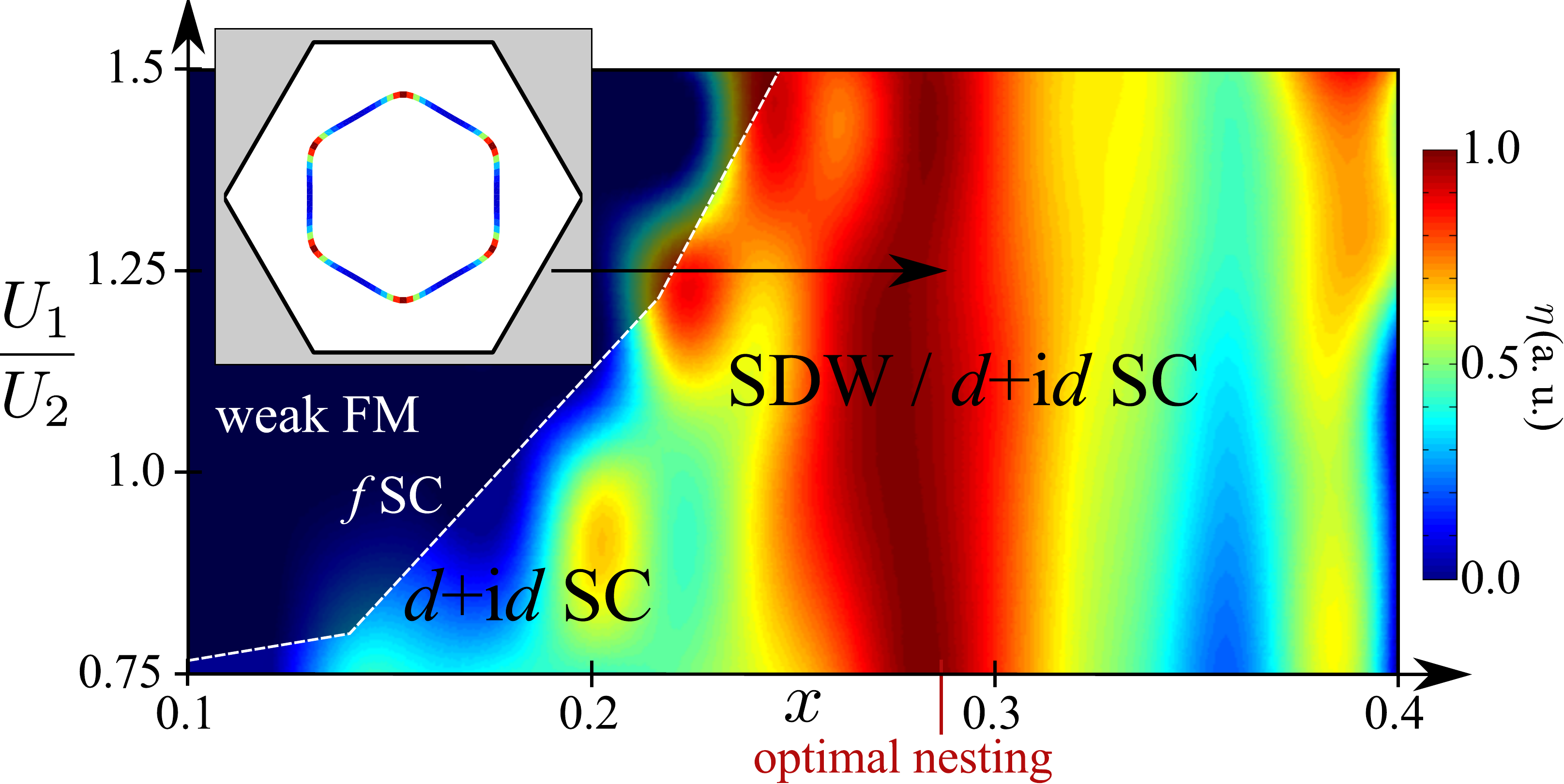}
    \end{minipage}
    \caption{(Color online). Gap anisotropy $\eta=\frac{\sigma(\Delta_0)}{\overline{\Delta_0}}$ in the $d+id$-wave phase (axis annotations as in Fig.~\ref{fig:phase_diagram}). Blue (red) regions indicate a rather homogeneous (anisotropic) $d$-wave gap. The inset shows the plot of the gap function on the Fermi surface at physically sensible doping:$\Delta_0$ exhibits close-to-nodal (blue) and larger (red) gap regions along the Fermi surface.}
    \label{fig:gap_map}
    \vspace{-0pt}
  \end{figure}

In Fig.~\ref{fig:gap_map}, we have plotted the degree of gap anisotropy in the $d+id$ phase for the same range of parameters as for the phase diagram in Fig.~\ref{fig:phase_diagram}. We use the variance of the gap function, divided by the mean: $\eta=\frac{\sigma(\Delta_0)}{\overline{\Delta_0}}$. Notably, the physically relevant regime for the cobaltates coincides with the strongest gap anisotropy, we observe. As seen in the inset of Fig.~\ref{fig:gap_map}, this yields a gap structure along a single Fermi surface which can be well characterized by a very small gap along the sides and a large gap at the edges of the Fermi surface. It can explain the two gap scenario from specific heat measurements~\cite{oeschler-08prb054528}, where both effective gap scales originate from a single pocket.

{\it Conclusion and perspectives.} We have shown that the experimental evidence of a close-to-nodal singlet superconducting state in the sodium cobaltates can be developed from a microscopic model taking into account the multi-orbital nature of the electronic scenario. The anisotropic $d+id$ superconductor we find is a combined effect of magnetic fluctuations, specific Fermi surface topology at the corresponding Na doping, and multi-orbital effects. To our conviction, this constitutes the sodium cobaltates to be one of the most promising candidates for a chiral singlet superconductor, to be further studied experimentally and theoretically. The T$_c$ might allow for laser-ARPES studies in the superconducting phase, along with more careful investigations of time-reversal symmetry breaking than it has been pursued by now. Likewise, the role of lattice distortions and disorder can  be interesting to consider, as there might be a transition from $d+id$ to $d$ when the custodial $C_{6v}$ symmetry is sufficiently broken. 
Finally, the desire for an ideal chiral singlet superconductor might also warrant further work to optimise its two-dimensional character, as has recently been reported for intercalated iron-based superconductors~\cite{burrard-lucas-13nm15}


\begin{acknowledgments}
RT thanks G.~Baskaran, B.~A.~Bernevig, S.~Blundell, A.~Boothroyd, S.~A.~Kivelson, and S.~Raghu for discussions. MK is supported by DFG-FOR 1162. CP, WH, and RT acknowledge support by SPP-1458.
\end{acknowledgments}


\end{document}